\documentclass{ws-procs9x6}
\newcommand{\open}{{<\kern -0.3 em{\scriptscriptstyle )}}}

\begin{document}

\title{Alternative approaches to transversity: \\
how convenient and feasible are they?}

\author{M. RADICI}

\address{Dipartimento di Fisica Nucleare e Teorica, Universit\`{a} di Pavia, and\\
Istituto Nazionale di Fisica Nucleare, Sezione di Pavia\\
via Bassi 6, I-27100 Pavia, Italy\\ 
E-mail: radici@pv.infn.it}

\maketitle

\abstracts{
The complete knowledge of the nucleon spin structure at leading twist 
requires also addressing the transverse spin distribution of quarks, or 
transversity, which is yet unexplored because of its chiral-odd nature. 
Elaborating strategies to extract it from (spin) asymmetry data represents a unique 
opportunity to explore more generally the transverse spin structure of the nucleon 
and the transverse momentum dynamics of partons inside it. Here, we critically 
review some of the most promising approaches.
}

\section{Introduction}
\label{sec:intro}

The predictions that the nucleon tensor charge is much larger than its helicity  and that the evolution of transversity should be weaker than the helicity one, are 
counterintuitive and they represent a basic test of QCD in the nonperturbative 
domain (for a review, see Refs.~\cite{Jaffe:1996zw,Barone:2003fy}). 

The pioneering suggestion of extracting the transversity from the Drell-Yan process 
with transversely polarized protons~\cite{Ralston:1979ys} opened the way to a deeper insight of the spin structure of the proton. In fact, if the cross section depends explicitly upon the transverse momentum of the lepton 
pair, interesting information can be inferred by using unpolarized and single-polarized Drell-Yan with antiproton beams. Two leading-twist convolutions of novel distribution functions open the door on studies of the orbital motion of 
partons inside hadrons~\cite{Boer:1999mm}. In Sec.~\ref{sec:dy}, we will simulate 
the corresponding spin asymmetries in order to explore the feasibility of such 
measurements at the future HESR facility at GSI. We will also compare the results 
with what can be expected if the antiproton beam is replaced by a pion beam in the
kinematic conditions reachable at COMPASS, such that the center-of-mass (cm) 
energy of the reaction is the same.

The growing interest in the transversity reflected in a rich experimental program 
at several laboratories. In particular, single-spin asymmetries have been measured in SIDIS with transversely polarized targets~\cite{Airapetian:2004tw}. A possible interpretation in terms of the Collins effect requires the cross section to depend explicitly upon the transverse momentum of the detected pion with respect to the jet axis. This fact brings in several complications, including the overlap with other competing mechanisms and more complicated factorization proofs and evolution equations.

It seems more convenient to consider more exclusive final states, where, e.g., 2 pions are inclusively detected~\cite{Collins:1994kq}. The chiral-odd partner of transversity can be represented by the so-called Interference Fragmentation Function 
(IFF) $H_1^{\open}$~\cite{Bianconi:1999cd} and it can be extracted by looking 
for asymmetric orientations of the plane containing the pion pair with
respect to the scattering plane~\cite{Radici:2001na}. In Sec.~\ref{sec:iff}, we 
will briefly recall the advantages of this strategy and present new results in comparison with upcoming data from the HERMES collaboration~\cite{NatDIS05}. 


\section{Single-polarized Drell-Yan at GSI and COMPASS}
\label{sec:dy}

The polarized part of the cross section for the process 
$\bar{p} p^\uparrow \rightarrow l^+ l^- X$ contains at leading 
twist the terms~\cite{Boer:1999mm}
\begin{eqnarray}
& &\frac{d\Delta \sigma^\uparrow}{d\Omega dx_1 dx_2 d{\bf q}_{_T}} \propto  
\sum_f\,e_f^2\,|{\bf S}_{2_T}|\,
\Bigg\{ - B(y) \, \sin  (\phi + \phi_{S_2})\, 
F \left[  \hat{\bf h}\cdot {\bf p}_{1_T} \,
\frac{\bar{h}_1^{\perp\,f}\, h_1^f}{M_1}\right]\nonumber \\
& &\mbox{\hspace{3cm}} + A(y) \, \sin (\phi - \phi_{S_2})\, 
F \left[ \hat{\bf h}\cdot {\bf p}_{2_T} \,
\frac{\bar{f}_1^f\, f_{1T}^{\perp\,f}}{M_2}\right] \; ... \Bigg\} \; ,
\label{eq:1polxsect}
\end{eqnarray}
where the annihilating partons with charge $e_f$ carry transverse momenta 
${\bf p}_{1,2_T}$ and longitudinal fractions $x_{1,2}$ of the proton momentum with mass $M$ and transverse polarization ${\bf S}_{2_T}$. The functions $A(y)$ and $B(y)$ depend only on the leptonic scattering angle via $y = (1+\cos \theta )/2$. The convolution $F$ is defined as
\begin{equation}
F \left[ \bar{f}_1  f_1 \right] = \int d{\bf p}_{1_T} d{\bf p}_{2_T}\, 
\delta ({\bf p}_{1_T}+{\bf p}_{2_T}-{\bf q}_{_T}) \left[ \bar{f}_1(x_1,{\bf p}_{1_T}) f_1(x_2,{\bf p}_{2_T})+1\leftrightarrow 2 \right] \, .
\label{eq:convol}
\end{equation}


\subsection{The Boer-Mulders effect}
\label{subsec:h1perp}

The first term in Eq.~(\ref{eq:1polxsect}) involves the transversity $h_1$ convoluted with the chiral-odd distribution $h_1^\perp$, which describes the influence of the quark transverse polarization on its momentum distribution inside an unpolarized parent hadron. Extraction of the latter is of great importance, because $h_1^\perp$ is believed to be responsible for the 
well known violation of the Lam-Tung sum rule~\cite{Conway:1989fs}, an anomalous big azimuthal asymmetry of the corresponding unpolarized Drell-Yan cross section  that still awaits for a justification.  

This contribution is simulated in a Monte Carlo along the lines described in 
Ref.~\cite{Bianconi:2004wu}. The spin asymmetry is produced by dividing the events into two groups, one for positive (U) and one for negative (D) values of $\sin (\phi + \phi_{S_2})$ for each bin $x_2$, and then constructing the
ratio $(U-D)/(U+D)$ for each bin $x_2$ after integrating upon $x_1, \theta ,$ and ${\bf q}_{_T}$. Two different test functions (ascending and descending) are used to probe the $x_2$ dependence. The goal is to explore under which conditions such different behaviours can be recognized also in the corresponding asymmetry $A_{_T}$. In fact, in that case the measurement of  $A_{_T}$ would allow for the extraction of unambigous information on the analytical form of both $h_1^\perp(x)$ and $h_1(x)$.

\begin{figure}[h]
\centerline{\epsfxsize=7cm\epsfbox{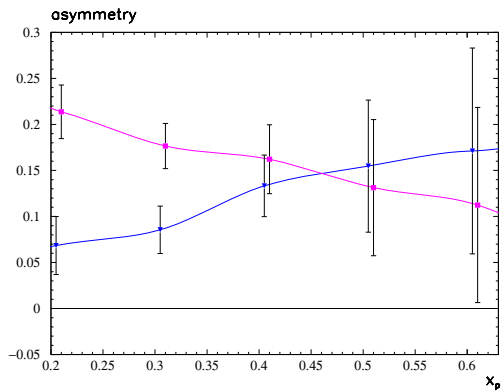}}   
\caption{Asymmetry $(U-D)/(U+D)$ from Boer-Mulders effect (see text) as 
function of $x_2$ for the process 
$\bar{p} p^\uparrow \rightarrow \mu^+ \mu^- X$. Full squares for the 
descending input test function, downward triangles for the ascending one (see text). Continuous lines are drawn to guide the eye. Error bars due to statistical errors only, obtained by 20 independent repetitions of the simulation.}
\label{fig:at-collider}
\end{figure}

In Fig.~\ref{fig:at-collider}, the asymmetry is shown for the 
$\bar{p} p^\uparrow \rightarrow \mu^+ \mu^- X$ process that could be realized at GSI in so-called asymmetric collider mode, i.e. with $E_{\bar{p}}=15$ GeV and $E_p=3.3$ GeV such that the cm energy squared is $s\approx 200$ (GeV)$^2$. The significant sample is made of 8000 selected events and statistical error bars are obtained by repeating 20 times the simulation for each bin $x_2$. For an hypothetical luminosity of $10^{31}$ cm$^{-2}$ s$^{-1}$, this corresponds to a running time of approximately three months (for further details see 
Ref.~\cite{Bianconi:2004wu}). From the figure, we deduce that in the range $0.1 < x_2 < 0.4$ it seems possible to extract information on the $x$ dependence of both the transversity and $h_1^\perp$. 

\begin{figure}[h]
\centerline{\epsfxsize=7cm\epsfbox{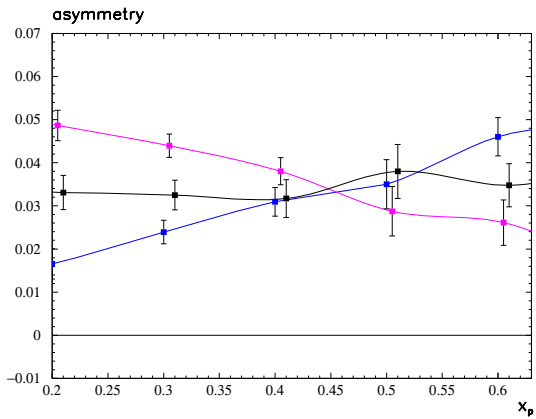}}   
\caption{Asymmetry $(U-D)/(U+D)$ (see text) as function of $x_2$ for the
process $\pi^- p^\uparrow \rightarrow \mu^+ \mu^- X$, using the same
conditions as in the previous figure. The case of constant input test function 
is also included. Continuous lines are drawn to guide the eye. Error bars due 
to statistical errors only, obtained by 10 independent repetitions of 
the simulation.}
\label{fig:at-compass}
\end{figure}

This statement can be reinforced if antiprotons are replaced by pions. The abundance of such particles allows for realistically increasing the significant sample by an order of magnitude. We selected the COMPASS setup with a $\pi^-$ beam of energy $E_\pi \sim 100$ GeV, impinging on a transversely polarized $NH_3$ fixed target. The corresponding $\pi^- p^\uparrow \rightarrow \mu^+ \mu^- X$ process at $s \sim 200$ (GeV)$^2$ can be directly compared with the asymmetric collider setup at GSI. Combining the increased statistics with the stronger dilution factor, for a sample of 125000 events we can considerably shrink the statistical error bars of the asymmetry, as it is evident from Fig.~\ref{fig:at-compass}.

\subsection{The Sivers effect}
\label{subsec:sivers}

The second term in Eq.~(\ref{eq:1polxsect}) has a different azimuthal dependence and it involves the standard unpolarized distribution $f_1$ and the Sivers function $f_{1T}^\perp$, which describes how the distribution of unpolarized quarks is affected by the transverse polarization of the parent proton. A measurement of a nonvanishing asymmetry would be a direct evidence of the orbital angular momentum of quarks. 

In order to perform the Monte Carlo simulation, we adopted for
$f_{_{1T}}^\perp(x,{\bf p}_{_T})$ the parametrization of 
Ref.~\cite{Anselmino:2005nn}, which was determined by fitting the HERMES data for one-pion inclusive production in DIS regime~\cite{Airapetian:2004tw} assuming that such asymmetry is produced by the Sivers effect only. We have further simplified the 
expression by neglecting the contribution of antipartons. Similarly to the Boer-Mulders case, the
asymmetry is generated in the Monte Carlo by dividing the events into two groups, one for positive (U) and one for negative (D) values of $\sin (\phi - \phi_{S_2})$ for each bin $x_2$, and then constructing the ratio $(U-D)/(U+D)$ for each bin $x_2$ after integrating upon $x_1, \theta ,$ and ${\bf q}_{_T}$.

\begin{figure}[h]
\centerline{\epsfxsize=7cm\epsfbox{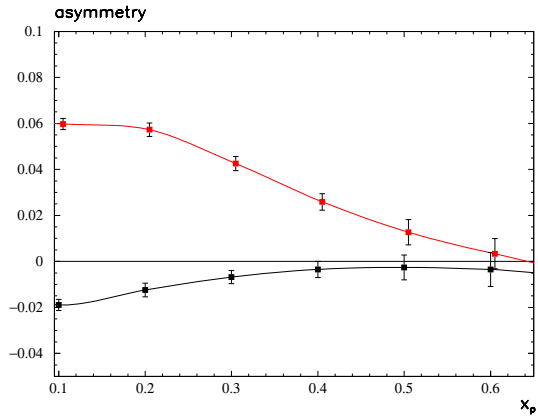}}   
\caption{Asymmetry $(U-D)/(U+D)$ from Sivers effect (see text) as function of $x_2$ for the process $\pi^\pm p^\uparrow \rightarrow \mu^+ \mu^- X$. Upper curve is the
$\pi^+$ case, lower curve the $\pi^-$ one. Continuous lines are drawn 
to guide the eye. Error bars due to statistical errors only, obtained 
by 10 independent repetitions of the simulation.}
\label{fig:atpi-sivers}
\end{figure}

In Fig.~\ref{fig:atpi-sivers}, the mechanism is considered for the process 
$\pi^\pm p^\uparrow \rightarrow \mu^+ \mu^- X$ in the COMPASS setup, as in 
Fig.~\ref{fig:at-compass}. The resulting statistical error bars are very small because of the high statistics, and selection of the beam charge can even determine the sign and size of the spin asymmetry.


\section{Two-hadron inclusive production at HERMES}
\label{sec:iff}

As already anticipated in Sec.~\ref{sec:intro}, looking for more exclusive final states in SIDIS can represent a competitive alternative to the Collins mechanism. In fact, for hadron pairs collinear with the jet axis the cross section at leading twist for two-hadron inclusive SIDIS production has a very simple structure: the unpolarized contribution and the factorized product of $h_1$ and $H_1^{\open}$,  
which describes the fragmentation of transversely polarized 
quarks~\cite{Radici:2001na}. There is no overlap with other mechanisms; collinear factorization holds and evolution properties of such objects should be determined straightforwardly~\cite{Boer:2001zw}. The unknown IFF can be extracted from the 
$e^+e^- \rightarrow (h_1 h_2)_{jet1} (h_1 h_2)_{jet2} X$ process by looking for an azimuthal asymmetry in the position of the hadron pair planes with respect to the lab frame~\cite{Boer:2003ya}. 

Since IFF are built on T-odd structures, it is possible to study in detail the mechanisms involved in the residual interactions of the outgoing hadrons by expanding the amplitudes in relative partial waves. If the hadrons are two pions, the main partial-wave contributions are the $s$ and $p$ 
waves~\cite{Jaffe:1998hf,Bacchetta:2002ux}. Residual interactions come from interference of amplitudes for different channels with different phases. Each interference component ($s-p$ or $p-p$) can be disentangled by a suitable selection of the integration phase space~\cite{Bacchetta:2002ux}. In particular, the following weighted asymmetry 
\begin{eqnarray}
A^{\sin (\phi_{_R}+\phi_{_S})\,\sin\theta}_{_{UT}} &= &
\frac{\displaystyle{\int} d\cos\theta \, d\phi_{_R} \, d\phi_{_S} \, 
       \sin (\phi_{_R}+\phi_{_S}) \, d\sigma_{_{UT}}}
     {\displaystyle{\int} d\cos\theta \, d\phi_{_R} \, d\phi_{_S} \, 
        d\sigma_{_{UU}}} \nonumber \\
&\propto &\frac{|{\bf S}_{_T}||{\bf R}|}{M_h} \, 
\frac{\sum_f e^2_f\, h_1^f\, H_{1,sp}^{\open \, f}}
     {\sum_f e^2_f\, f_1^f\, (\frac{3}{4} \, D_1^{pp\,f} +
       \frac{1}{4}\, D_1^{ss\,f})} 
\label{eq:2hssaLM}
\end{eqnarray}
has been measured at HERMES for the $e p^\uparrow \rightarrow e' (\pi \pi) X$ process, where the proton polarization ${\bf S}_{_T}$ forms an azimuthal angle $\phi_{_S}$ with the scattering plane, and the final pion pair with invariant mass $M_h$ has a relative momentum ${\bf R}$ oriented with the azimuthal angle $\phi_{_R}$. The angle $\theta$ defines the direction of the back-to-back emission in the pair cm frame with respect to the jet axis. 

In the literature, there are two predictions for the asymmetry of 
Eq.~(\ref{eq:2hssaLM}). The first one~\cite{Jaffe:1998hf} is based on the guess that the asymmetry should depend on the properties of the $\pi - \pi$ phase shifts for the considered $s$ and $p$ channels. The second one~\cite{Radici:2001na} is based on the calculation in the spectator model of the interference diagram where the two pions are emitted directly or through the decay of the $\rho$ resonance. The results have strikingly different features, because the first model predicts a marked $M_h$ dependence with a sign change around $M_h \approx m_\rho$, which is not observed in the second model. 

Therefore, we have considered a more refined version of the spectator 
model~\cite{noiIFFnuovo}. The amplitude for the $p$ channel contains the coherent sum of the
resonant decays $\rho, \, \omega \rightarrow \pi^+ \pi^-$, and the incoherent sum of the
channel $\omega \rightarrow \pi^+ \pi^- \pi^0$, properly integrated upon the third pion
$\pi^0$. In the $s$ channel, the amplitude is the incoherent sum of the $K_s^0 \rightarrow
\pi^+ \pi^-$ decay and of a background, represented by the direct production of the charged
pion pair. The diagonal $s$ and $p$ contributions enter $D_1^{ss}$ and $D_1^{pp}$ in
Eq.~(\ref{eq:2hssaLM}), respectively, while $H_{1\, sp}^{\open}$ contains the $s-p$ interference. 

The parameters of the resonances are taken from the Particle Data 
Group~\cite{Eidelman:2004wy}, while the free parameters of the model are then fixed by reproducing the invariant mass distribution $D_1(M_h)$ as it is output by the HERMES Monte Carlo program with no corrections for acceptance and with proper kinematical cuts in order to remove elastic, single- and double-diffractive events~\cite{elke}.

\begin{figure}[h]
\centerline{\epsfxsize=7cm\epsfbox{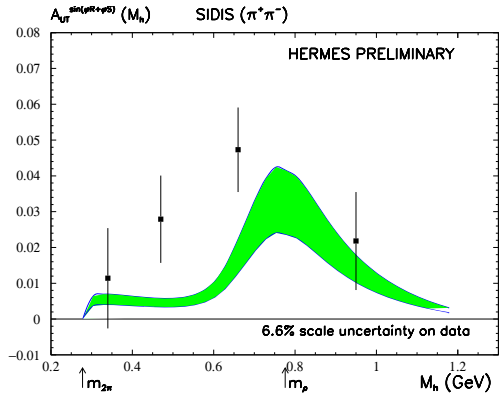}}  
\caption{Preliminary HERMES data and results for the single spin asymmetry from the 
interference of 2 pion production channels in relative $s$ and $p$ waves, for the 
process $e p^\uparrow \rightarrow e' (\pi \pi) X$.}
\label{fig:aut}
\end{figure}

Once all the free parameters have been fixed to the invariant mass 
spectrum, we predict the single-spin asymmetry 
$A_{_{UT}}^{\sin (\phi_{_R}+\phi_{_S}) \, \sin\theta}$ of Eq.~(\ref{eq:2hssaLM}). We further integrate it upon all variables but $M_h$, following the experimental cuts. The net result is compared in Fig.~\ref{fig:aut} with available preliminary experimental data~\cite{NatDIS05}; in particular, no evidence for a sign change of the asymmetry is displayed. The uncertainty band is given only by various possible choices for the distribution functions $f_1, h_1$. All the other ingredients of the calculations are fixed and, therefore, we can interpret the result as a true prediction.  


\section*{Acknowledgments}
A fruitful collaboration is warmly acknowledged with A.~Bacchetta and A.~Bianconi, who coauthored most of the results here presented.


\end{document}